\documentclass[twocolumn,showpacs,preprintnumbers,amsmath,amssymb]{revtex4}
\usepackage{graphicx}
\usepackage{dcolumn}
\usepackage{bm}

\begin{document}

\title{Pygmy dipole mode in deformed neutron-rich Mg isotopes close to the drip line
}

\author{Kenichi Yoshida}
\affiliation{RIKEN Nishina Center for Accelerator-Based Science, Wako, Saitama 351-0198, Japan
}%

\date{\today}

\begin{abstract}
We investigate the microscopic structure of 
the low-lying isovector-dipole excitation mode 
in neutron-rich $^{36,38,40}$Mg close to the drip line 
by means of 
the deformed quasiparticle-random-phase-approximation 
employing the Skyrme and the local pairing energy-density functionals. 
It is found that the low-lying bump structure above the neutron emission-threshold energy 
develops when the drip line is approached, and 
that the isovector dipole strength at $E_{x} < 10$ MeV exhausts about $6.0 \%$ of the classical 
Thomas-Reiche-Kuhn dipole sum rule in $^{40}$Mg. 
We obtained the collective dipole modes at around $8-10$ MeV in Mg isotopes, 
which consist of many two-quasiparticle excitations of neutron. 
The transition density clearly shows an oscillation of the neutron skin against 
the isoscalar core. 
We found significant coupling effects 
between the dipole and octupole excitation modes due to the nuclear deformation. 
It is also found that the responses for the compressional dipole and isoscalar octupole 
excitations are much enhanced in the lower energy region. 
\end{abstract}

\maketitle

\section{Introduction}
Nuclei far from the stability has attracted a considerable interest 
both experimentally and theoretically. 
Exploring the multipole responses in unstable nuclei 
is in particular of great interest 
because they provide information on 
collective modes of excitation. 
In neutron-rich nuclei, 
the surface structure is quite different to the stable ones 
due to the presence of the loosely bound neutrons. 
One of the unique structures is the neutron skin~\cite{suz95,miz00}. 
Since the collective excitations are sensitive to the surface structure, 
we can expect new kinds of exotic excitation mode associated with the neutron skin 
to appear in neutron-rich nuclei. 
An example is the soft dipole excitation~\cite{ike88}, 
or the pygmy dipole resonance (PDR), 
which is observed not only 
in light halo nuclei~\cite{sac93,shi95,zin97,nak06,nak94,pal03,fuk04,nak99,pra03,aum99}, 
but also in heavier systems~\cite{lei01,try03,gib08,adr05}, where an appreciable 
$E1$ strength is observed above the neutron emission threshold, and 
it exhausts several percents of the energy-weighted sum rule (EWSR) value. 

The structure of the PDR and its collectivity 
has been studied based on the mean-field calculations 
by many groups. (See Ref.~\cite{paa07} for 
extensive lists of references concerning the theoretical investigation of the PDR.)
These studies, however, are largely restricted to spherical systems.
Quite recently, by means of the deformed quasiparticle-random-phase approximation (QRPA), 
the low-lying dipole excitation in neutron-rich Ne isotopes~\cite{yos08a} 
and in neutron-rich Sn isotopes~\cite{pen09} have been investigated. 

Presently, small excitation energies of the first $2^{+}$ state and 
striking enhancements of $B(E2;0_{1}^{+} \to 2_{1}^{+})$ 
in $^{32}$Mg~\cite{gui84,mot95} and $^{34}$Mg~\cite{yon01,chu05,ele06} 
are under lively discussions 
in connection with onset of the quadrupole deformation, 
breaking of the $N=20$ spherical magic number, 
pairing correlation and continuum coupling effects~\cite{pov87,war90,fuk92,uts99,yam04}. 
In order to get clear understanding of the nature of quadrupole deformation 
and pairing correlations, 
it is strongly desirable to explore, both experimentally and theoretically, 
excitation modes in Mg isotopes 
toward a drip line~\cite{gad07,bau07,ter97,cau98,rei99,rod02,yos06}. 

In the present article, we investigate the microscopic structure of the low-lying 
dipole excitation in neutron-rich Mg isotopes close to the drip line, 
paying special attention to the deformation effects on them. 
In a deformed system, the soft negative-parity excitation modes 
could emerge associated with coupling between the dipole and octupole modes 
of excitation~\cite{ina02}. 
To this end, 
we perform a deformed QRPA in the matrix formalism 
on top of the coordinate-space Skyrme-Hartree-Fock-Bogoliubov (HFB) theory 
developed in Ref.~\cite{yos08}. 
The matrix formalism of the QRPA is suitable to investigate 
the microscopic structure of the excitation modes. 

This article is organized as follows: 
In the next section, the deformed Skyrme-HFB + QRPA method is recapitulated. 
In Sec.~\ref{results}, we show the results of the deformed QRPA and investigate 
microscopic structures of the low-lying dipole states in $^{36,38,40}$Mg. 
We discuss properties of the coupling among different modes of excitation 
unique in deformed neutron-rich nuclei in Sec.~\ref{discussion}. 
The last section is devoted to a summary.

\section{\label{model}Model}
A detailed discussion of the deformed Skyrme-HFB + QRPA can be found in Ref.~\cite{yos08}. 
Therefore, we just briefly recall the outline of the calculation scheme. 

In order to describe simultaneously the nuclear deformation 
and the pairing correlations including the unbound quasiparticle states, 
we solve the HFB equations~\cite{dob84,bul80}
\begin{equation}
\begin{pmatrix}
h^{q}-\lambda^{q} & \tilde{h}^{q} \\
\tilde{h}^{q} & -(h^{q}-\lambda^{q})
\end{pmatrix}
\begin{pmatrix}
\varphi^{q}_{1,\alpha}(\boldsymbol{r},\sigma) \\
\varphi^{q}_{2,\alpha}(\boldsymbol{r},\sigma)
\end{pmatrix} 
= E_{\alpha}
\begin{pmatrix}
\varphi^{q}_{1,\alpha}(\boldsymbol{r},\sigma) \\
\varphi^{q}_{2,\alpha}(\boldsymbol{r},\sigma)
\end{pmatrix} \label{HFB_equation}
\end{equation}
in coordinate space using cylindrical coordinates $\boldsymbol{r}=(\rho,z,\phi)$.
We assume axial and reflection symmetries.
Here, $q=\nu$ (neutron) or $\pi$ (proton).
For the mean-field Hamiltonian $h$, we employ the SkM* interaction~\cite{bar82}. 
Details for expressing the densities and currents in the cylindrical coordinate
representation can be found in Ref.~\cite{ter03}.
The pairing field is treated by using the density-dependent contact
interaction~\cite{cha76},
\begin{equation}
v_{pair}(\boldsymbol{r},\boldsymbol{r}^{\prime})=\dfrac{1-P_{\sigma}}{2}
\left[ t_{0}^{\prime}+\dfrac{t_{3}^{\prime}}{6}\varrho_{0}^{\gamma}(\boldsymbol{r}) \right]
\delta(\boldsymbol{r}-\boldsymbol{r}^{\prime}). \label{pair_int}
\end{equation}
where $\varrho_{0}(\boldsymbol{r})$ denotes the isoscalar density of the ground state 
and $P_{\sigma}$ the spin exchange operator.
Assuming time-reversal symmetry and reflection symmetry with respect to the $x-y$ plane,
we have to solve for positive $\Omega$ and positive $z$ only, 
$\Omega$ being the $z-$component of the angular momentum $j$. 
We use the lattice mesh size $\Delta\rho=\Delta z=0.6$ fm and a box
boundary condition at $\rho_{\mathrm{max}}=9.9$ fm, $z_{\mathrm{max}}=12$ fm. 
The differential operators are represented by use of the 11-point formula of Finite Difference Method. 
Because the parity and $\Omega$ are good quantum numbers in the present calculation scheme, 
we have only to diagonalize the HFB Hamiltonian (\ref{HFB_equation}) for each $\Omega^{\pi}$ sector. 
The quasiparticle energy is cut off at $E_{\mathrm{qp,cut}}=60$ MeV
and the quasiparticle states up to $\Omega^{\pi}=15/2^{\pm}$ are included.

The pairing strength parameter $t_{0}^{\prime}$ is
determined so as to reproduce the experimental pairing gap of
$^{34}$Mg ($\Delta_{\mathrm{exp}}=1.7$ MeV) obtained by the
three-point formula~\cite{sat98}.
The strength $t_{0}^{\prime}=-295$ MeV fm$^{3}$ for the
mixed-type interaction ($t_{3}^{\prime}=-18.75t_{0}^{\prime}$)~\cite{ben05} 
with $\gamma=1$ leads to the pairing gap
$\langle \Delta_{\nu}\rangle=1.71$ MeV in $^{34}$Mg.

Using the quasiparticle basis obtained
as a self-consistent solution of the HFB equations (\ref{HFB_equation}),
we solve the QRPA equation in the matrix formulation~\cite{row70}
\begin{equation}
\sum_{\gamma \delta}
\begin{pmatrix}
A_{\alpha \beta \gamma \delta} & B_{\alpha \beta \gamma \delta} \\
-B_{\alpha \beta \gamma \delta} & -A_{\alpha \beta \gamma \delta}
\end{pmatrix}
\begin{pmatrix}
X_{\gamma \delta}^{i} \\ Y_{\gamma \delta}^{i}
\end{pmatrix}
=\hbar \omega_{i}
\begin{pmatrix}
X_{\alpha \beta}^{i} \\ Y_{\alpha \beta}^{i}
\end{pmatrix} \label{eq:AB1}.
\end{equation}
The residual interaction in the particle-hole (p-h) channel appearing
in the QRPA matrices $A$ and $B$ is
derived from the Skyrme density functional. 
We neglect the spin-orbit interaction term
$C_{t}^{\nabla J}$ as well as the
Coulomb interaction to reduce the computing time in the QRPA calculation.
We also drop the so-called $``{J}^{2}"$ term $C_{t}^{T}$ both in 
the HFB and QRPA calculations. 
The residual interaction in the
particle-particle (p-p) channel is derived from the pairing
functional constructed with the density-dependent contact
interaction (\ref{pair_int}).

Because the full self-consistency between the static mean-field
calculation and the dynamical calculation is broken by the above
neglected terms, we renormalize the residual interaction in the
p-h channel by an overall factor $f_{\mathrm{ph}}$ to get the spurious mode. 
We cut the two-quasiparticle
(2qp) space at $E_{\alpha}+E_{\beta} \leq 60$ MeV due to the
excessively demanding computer memory size and computing time
for the model space consistent with that adopted in the HFB
calculation; $2 E_{\mathrm{qp,cut}}=120$ MeV.
Accordingly, we need another factor $f_{\mathrm{pp}}$ for the p-p channel. 
See Ref.~\cite{yos08} for details of determination of the normalization factors.
In the present calculation, 
the dimension of the QRPA matrix (\ref{eq:AB1}) 
for the $K^{\pi}=0^{-}$ excitation in $^{40}$Mg 
is about 17 100, and the memory size is 24.4 GB. 
The normalization factors are $f_{\mathrm{ph}}=1.06$, and $f_{\mathrm{pp}}=1.21$. 

\section{\label{results}Results of the calculation}
We summarize in Table~\ref{GS} the ground state properties.
The neutron-rich Mg isotopes under investigation are prolately deformed. 
This is consistent with the results calculated 
using the Skyrme SIII interaction~\cite{ter97}. 
The Gogny-HFB calculation using the D1S interaction 
suggested the shape coexistence in $^{38,40}$Mg~\cite{rod02}. 
We can see that the neutron skin develops as approaching the drip line; 
the difference in neutron and proton radii 
$\sqrt{\langle r^{2} \rangle_{\nu}}-\sqrt{\langle r^{2} \rangle_{\pi}}=0.41$ fm 
in $^{36}$Mg changes to 0.54 fm in $^{40}$Mg. 

\begin{table}[t]
\begin{center}
\caption{Ground state properties of $^{36,38,40}$Mg obtained by the deformed HFB calculation 
with the SkM* interaction and the mixed-type pairing interaction. Chemical potentials, 
deformation parameters, average pairing gaps, root-mean-square radii for 
neutrons and protons are listed. The average pairing gaps of protons are zero in these isotopes. }
\label{GS}
\begin{tabular}{cccc}
\hline \hline
\noalign{\smallskip}
  & $^{36}$Mg & $^{38}$Mg & $^{40}$Mg  \\
\noalign{\smallskip}\hline\noalign{\smallskip}
$\lambda_{\nu}$ (MeV)  & $-3.24$ & $-2.41$ & $-1.56$ \\
$\lambda_{\pi}$ (MeV)  & $-21.0$ & $-23.7$ & $-24.4$ \\
$\beta_{2}^{\nu}$  & 0.31 & 0.29 & 0.28 \\
$\beta_{2}^{\pi}$  & 0.39 & 0.38 & 0.36 \\
$\langle \Delta \rangle_{\nu}$ (MeV)  & 1.71 & 1.64 & 1.49 \\
$\sqrt{\langle r^{2} \rangle_{\nu}}$ (fm)  & 3.59 & 3.67 & 3.76 \\
$\sqrt{\langle r^{2} \rangle_{\pi}}$ (fm)  & 3.18 & 3.20 & 3.22 \\
\noalign{\smallskip}
\hline \hline
\end{tabular}
\end{center}
\end{table}

\begin{figure*}[t]
\begin{center}
\includegraphics[scale=0.9]{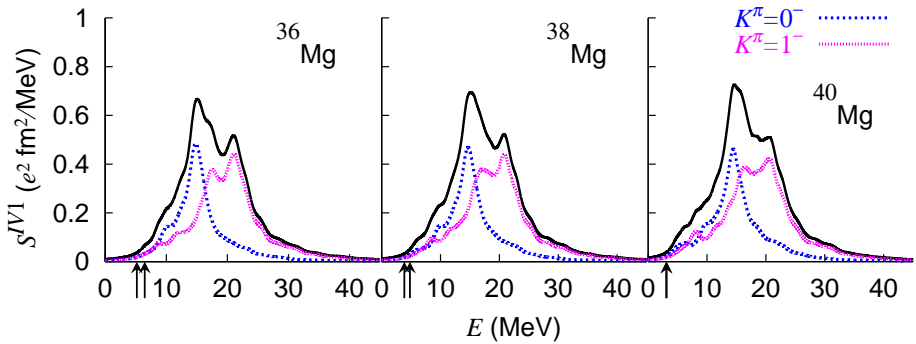}
\caption{(Color online) 
Response functions for the isovector (IV) dipole operator (\ref{IV1}) in $^{36,38,40}$Mg. 
The dotted, dashed and solid lines correspond to the $K^{\pi}=0^{-}$, $K^{\pi}=1^{-}$ and 
total responses, respectively. 
For the $K^{\pi}=1^{-}$ response, the transition strengths for the $K^{\pi}=\pm1^{-}$ states 
are summed up. The transition strengths are smeared by using  $\Gamma=2$ MeV. 
The arrows indicate the one-neutron continuum threshold $E_{\mathrm{th},1n}=|\lambda|+\min E_{\alpha}$ 
and the two-neutron continuum threshold $E_{\mathrm{th},2n}=2|\lambda|$. 
In $^{40}$Mg, these two continuum-threshold energies are almost degenerated. 
}
\label{response}
\end{center}
\end{figure*}

Figure~\ref{response} shows the response functions for the isovector (IV) dipole excitation in 
neutron-rich Mg isotopes. 
The IV dipole operator used in the present calculation is
\begin{equation}
\hat{F}^{\mathrm{IV}}_{1K}=e\dfrac{N}{A}\sum_{i \in \pi}r_{i}Y_{1K}(\hat{r}_{i})-
e\dfrac{Z}{A}\sum_{i \in \nu}r_{i}Y_{1K}(\hat{r}_{i}), \label{IV1}
\end{equation}
and the response function is calculated as
\begin{equation}
S^{\tau \lambda}(E)=\sum_{i}\sum_{K} \dfrac{\Gamma/2}{\pi}
\dfrac{|\langle i|\hat{F}^{\tau}_{\lambda K}|0\rangle|^{2}}
{(E-\hbar \omega_{i})^{2}+\Gamma^{2}/4}.
\end{equation} 

The Giant Dipole Resonance (GDR) appearing at $15-25$ MeV shows a deformation splitting 
for the $K^{\pi}=0^{-}$ and $1^{-}$ excitations. 
In the lower energy region, we can see a bump structure above the neutron-emission threshold 
energy.

\begin{figure}[t]
\begin{center}
\includegraphics[scale=0.55]{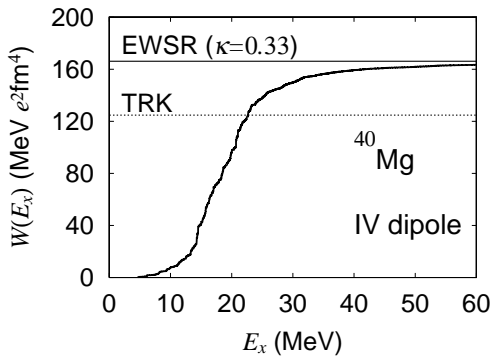}
\caption{Energy weighted sum of the IV dipole strength function in $^{40}$Mg. 
The horizontal lines show the classical Thomas-Reiche-Kuhn (TRK), and
the RPA sum rule values including the enhancement factor,
$m_{1}=m_{1}^{\mathrm{cl}}(1+\kappa)$ ($\kappa=0.33$ in $^{40}$Mg
with the SkM* interaction ). }
\label{40Mg_EWSR}
\end{center}
\end{figure}

Figure~\ref{40Mg_EWSR} shows the partial sum of the energy weighted
strength defined as
\begin{equation}
W(E_{x})=\sum_{\hbar \omega_{i}<E_{x}} \sum_{K} \hbar \omega_{i}
|\langle i|\hat{F}_{\lambda K}^{\tau}|0 \rangle|^{2}.
\end{equation}
For the IV dipole mode in $^{40}$Mg, the calculated sum up to 60 MeV
reaches 98.5\% of the EWSR value including the enhancement factor,
$m_{1}=m_{1}^{\mathrm{cl}}(1+\kappa)$ where
$\kappa=0.33$. 
The IV dipole strength below 10 MeV exhausts about $6.0 \%$ of 
the classical Thomas-Reiche-Kuhn (TRK) sum rule. 
In $^{36}$Mg and $^{38}$Mg, 
the summed transition strength up to 10 MeV 
exhausts about $3.6 \%$ and $4.8 \%$ of the TRK sum rule, respectively.

In what follows, the low-energy dipole excitations are 
investigated in detail.

\begin{figure*}[t]
\begin{center}
\includegraphics[scale=0.75]{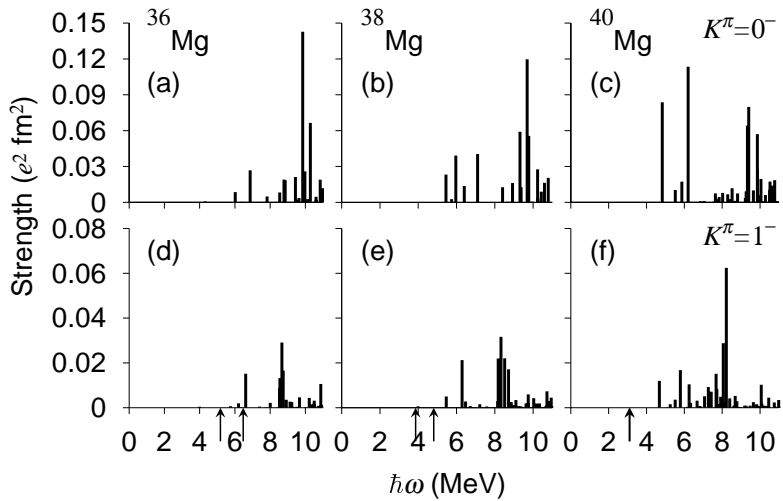}
\caption{
IV dipole transition strengths in $^{36,38,40}$Mg for 
the $K^{\pi}=0^{-}$ (upper) and $1^{-}$ (lower) excitations.}
\label{Mg_dip_strength}
\end{center}
\end{figure*}

\subsection{Low lying states in $^{40}$Mg}

\begin{table}[b]
\caption{QRPA amplitudes for the $K^{\pi}=1^{-}$ state in $^{40}$Mg at 8.22 MeV.
This mode has the isovector (IV) dipole strength 
$B(\mathrm{IV}1)=6.24 \times10^{-2} e^{2}$fm$^{2}$, 
the isoscalar (IS) octupole strength $B(\mathrm{IS}3)=1.44 \times 10^{3}$ fm$^{6}$, 
and the sum of backward-going amplitude $\sum Y_{\alpha\beta}^{2}=5.56\times 10^{-3}$. 
The single-quasiparticle levels are labeled with 
the asymptotic quantum numbers $[Nn_{3}\Lambda]\Omega$. 
Only components with $X_{\alpha\beta}^{2}-Y_{\alpha\beta}^{2} > 0.02$ are listed. 
Two-quasiparticle excitation energies are given by $E_{\alpha}+E_{\beta}$ in MeV and 
two-quasiparticle dipole transition matrix elements $D_{1,\alpha\beta}$ in $e$ fm, 
the octupole transition matrix elements $O_{1,\alpha\beta}$ in fm$^{3}$. 
In the rows (e) and (j), the label 
$\nu 3/2^{+}$ denotes a non-resonant discretized continuum state of neutron 
$\Omega^{\pi}=3/2^{+}$ level. 
The quasiparticle resonance of the hole-like $\nu[200]1/2$ level 
is described by three discretized states in the present box size, 
and the level in the row (b) is the same state as in (g).
}
\label{40Mg_1-}
\begin{center} 
\begin{tabular}{ccccccc}
\hline \hline
\noalign{\smallskip}
 & $\alpha$ & $\beta$  & $E_{\alpha}+E_{\beta}$ & $X_{\alpha \beta}^{2}-Y_{\alpha\beta}^{2}$  & 
$D_{1,\alpha\beta}$ & $O_{1,\alpha\beta}$  \\
 &  &  & (MeV) &  
 & ($e$ fm) & (fm$^{3}$) 
\\ \noalign{\smallskip} \hline \noalign{\smallskip}
(a) & $\nu[202]3/2$ & $\nu[321]1/2$ & 8.28 & 0.289 & $0.164$ & $-14.8$ \\ 
(b) & $\nu[200]1/2$ & $\nu[312]3/2$ & 8.26 & 0.167 & $-0.190$ & $10.5$ \\
(c) & $\nu[321]3/2$ & $\nu[440]1/2$ & 7.49 & 0.062 & $0.054$ & $29.7$ \\
(d) & $\nu[303]7/2$ & $\nu[422]5/2$ & 7.92 & 0.050 & $0.059$ & $3.82$ \\ 
(e) & $\nu[310]1/2$ & $\nu 3/2^{+}$ & 7.96 & 0.037 & $0.159$ & $16.4$ \\
(f) & $\nu[200]1/2$ & $\nu[312]3/2$ & 9.04 & 0.034 & $0.011$ & $-5.05$ \\
(g) & $\nu[200]1/2$ & $\nu[310]1/2$ & 7.72 & 0.030 & $-0.053$ & $-5.72$ \\ 
(h) & $\nu[303]7/2$ & $\nu[413]5/2$ & 9.74 & 0.021 & $-0.087$ & $10.7$ \\  
(i) & $\nu[312]5/2$ & $\nu[411]3/2$ & 6.91 & 0.020 & $0.073$ & $23.4$ \\
(j) & $\nu[312]5/2$ & $\nu 3/2^{+}$ & 8.97 & 0.020 & $0.106$ & $-2.29$ \\
\noalign{\smallskip}
\hline \hline
\end{tabular}
\end{center} 
\end{table}

Due to the deformation, the strength distribution and microscopic structure 
of the $K^{\pi}=0^{-}$ and $1^{-}$ excitations are different. 
Figures~$\ref{Mg_dip_strength}$(c) and $\ref{Mg_dip_strength}$(f) show the IV dipole strengths 
in the lower energy region for the $K^{\pi}=0^{-}$ and $1^{-}$ excitations in $^{40}$Mg. 

First, we are going to discuss the structure of the $K^{\pi}=1^{-}$ excitations. 
At $\hbar \omega = 8.22$ MeV, we can see a prominent peak possessing the large transition strength 
in Fig.~\ref{Mg_dip_strength}(f). 
We made a detailed analysis of this eigenmode and show in Table~\ref{40Mg_1-} 
its microscopic structure. 
This state is generated by many 2qp excitations. 
Among the 2qp excitations listed in Table~\ref{40Mg_1-}, 
the 2qp excitation of (d) and (h) is the particle-particle like excitation, 
and that of (g) is the hole-hole like excitation. 
These 2qp excitations never participate in generating the RPA mode 
in the absence of the pairing correlations.

\begin{figure}[t]
\begin{center}
\begin{tabular}{cc}
\includegraphics[scale=0.5]{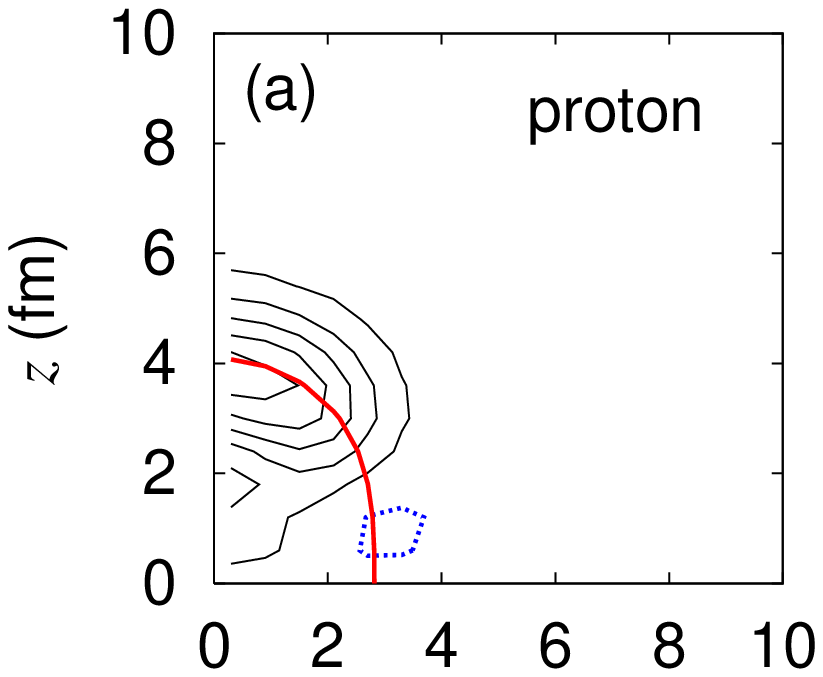}
\includegraphics[scale=0.5]{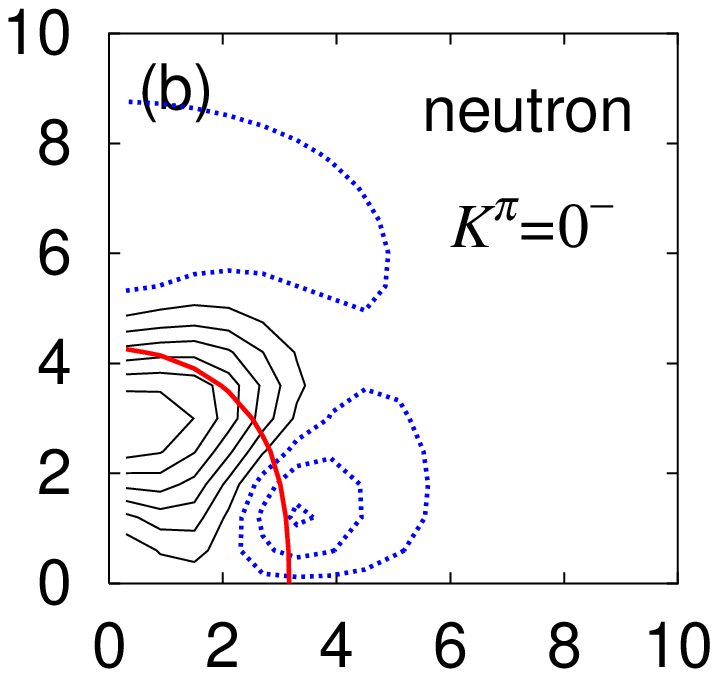} \\
\includegraphics[scale=0.5]{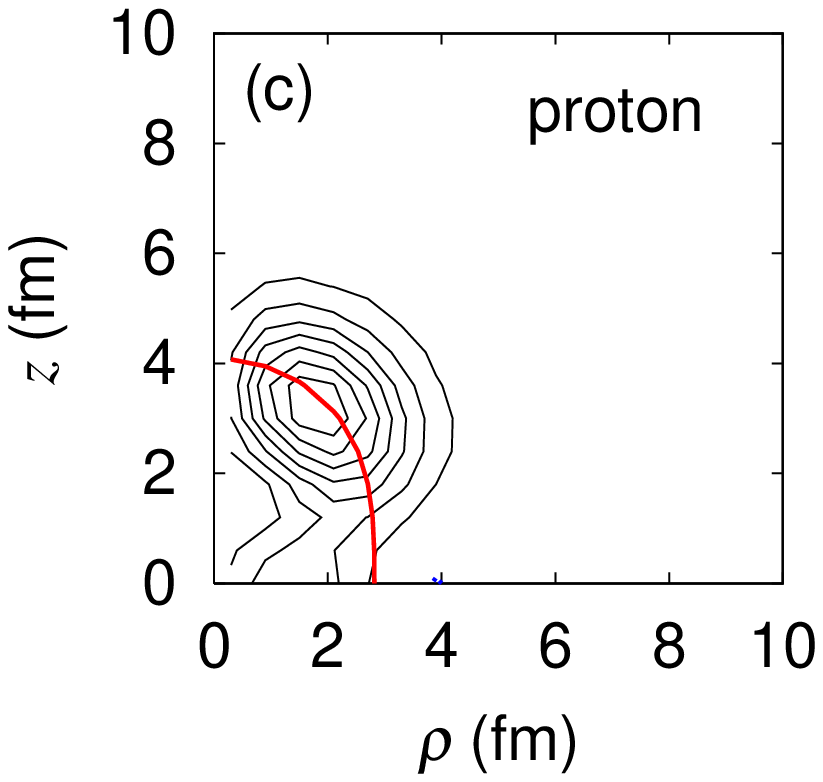}
\includegraphics[scale=0.5]{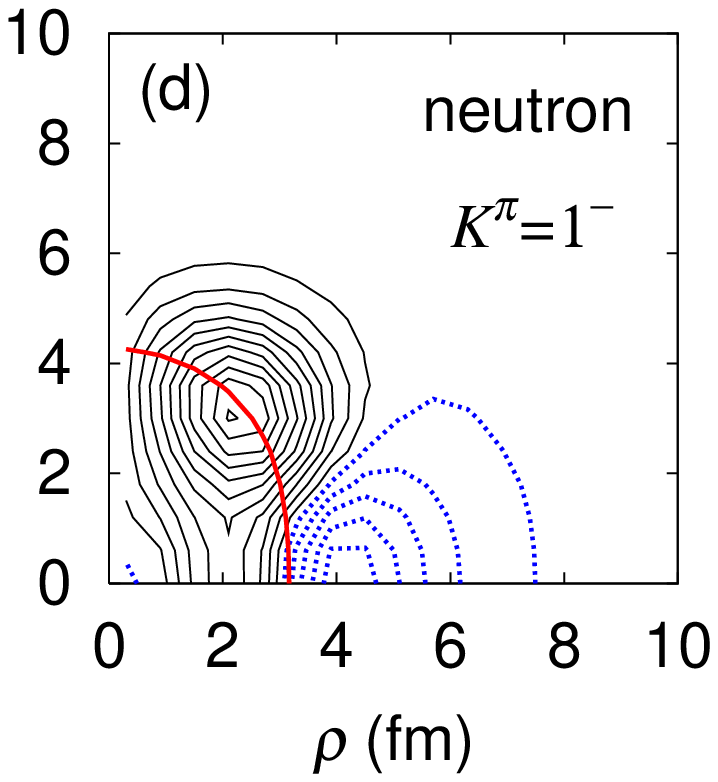}
\end{tabular}
\caption{(Color online) Transition densities of protons and neutrons in $^{40}$Mg 
to the $K^{\pi}=0^{-}$ state at 9.40 MeV (upper) and the $K^{\pi}=1^{-}$ state at 8.22 MeV (lower). 
Solid and dotted lines indicate positive 
and negative transition densities, and 
the contour lines are plotted at intervals of $3 \times 10^{-4}$ fm$^{-3}$. 
The thick solid lines indicate the neutron and proton half density, 
0.055 fm$^{-3}$ and 0.032 fm$^{-3}$, respectively.
}
\label{trans_density}
\end{center}
\end{figure}

In order to understand the spatial structure of the state at 8.22 MeV, 
we show in the lower panel of Fig.~\ref{trans_density} the transition density to this state. 
The transition density has the following features: 
Around the surface and the inside of the nucleus, 
the protons and neutrons oscillate with coherence. 
Outside of the nucleus, the neutrons only oscillate and 
the oscillation of the internal core and the neutron outside 
is in out-of-phase. 
This corresponds to the classical picture of 
an oscillation of the neutron skin against 
the isoscalar core along $\rho-$axis; the axis perpendicular to the symmetry axis.

\begin{table}[t]
\caption{
Same as Table~\ref{40Mg_1-} but for the $K^{\pi}=0^{-}$ state at 9.40 MeV
This mode has the IV dipole strength 
$B(\mathrm{IV}1)=7.99 \times10^{-2} e^{2}$fm$^{2}$, 
the IS octupole strength $B(\mathrm{IS}3)=23.8$ fm$^{6}$, 
and the sum of backward-going amplitude $\sum Y_{\alpha\beta}^{2}=8.76\times 10^{-4}$. 
In the rows (b), (d), (f) and (i), the label 
$\nu 1/2^{+}$ and $\nu 3/2^{+}$ denote 
non-resonant discretized continuum states of neutron 
$\Omega^{\pi}=1/2^{+}$ and $3/2^{+}$ levels. 
The quasiparticle resonance of the hole-like $\nu[330]1/2$ level 
is described by two discretized states in the present box size, 
and the level in the row (g) is the same state as in (i).
}
\label{40Mg_0-}
\begin{center} 
\begin{tabular}{ccccccc}
\hline \hline
\noalign{\smallskip}
 & $\alpha$ & $\beta$  & $E_{\alpha}+E_{\beta}$ & $X_{\alpha \beta}^{2}-Y_{\alpha\beta}^{2}$  & 
$D_{0,\alpha\beta}$ & $O_{0,\alpha\beta}$  \\
 &  &  & (MeV) &  
 & ($e$ fm) & (fm$^{3}$) 
\\ \noalign{\smallskip} \hline \noalign{\smallskip}
(a) & $\nu[200]1/2$ & $\nu[301]1/2$ & 9.34 & 0.376 & $0.058$ & $-3.31$ \\ 
(b) & $\nu[310]1/2$ & $\nu 1/2^{+}$ & 9.67 & 0.151 & $0.161$ & $-3.35$ \\
(c) & $\nu[330]1/2$ & $\nu[440]1/2$ & 8.85 & 0.111 & $0.256$ & $27.1$ \\
(d) & $\nu[321]1/2$ & $\nu 1/2^{+}$ & 9.37 & 0.072 & $0.011$ & $2.96$ \\ 
(e) & $\nu[321]3/2$ & $\nu[411]3/2$ & 8.64 & 0.056 & $0.311$ & $9.55$ \\
(f) & $\nu[312]3/2$ & $\nu 3/2^{+}$ & 9.36 & 0.052 & $-0.010$ & $3.52$ \\
(g) & $\nu[330]1/2$ & $\nu[440]1/2$ & 8.55 & 0.031 & $0.237$ & $29.6$ \\
(h) & $\nu[312]5/2$ & $\nu[413]5/2$ & 10.3 & 0.027 & $0.155$ & $7.69$ \\ 
(i) & $\nu[330]1/2$ & $\nu 1/2^{+}$ & 10.1 & 0.024 & $0.128$ & $6.89$ \\
\noalign{\smallskip}
\hline \hline
\end{tabular}
\end{center} 
\end{table}

We are going to move on to the $K^{\pi}=0^{-}$ excitations. 
Above the threshold energy, we can see several states possessing enhanced strengths in 
Fig.~\ref{Mg_dip_strength}(c).  
The states at 4.83 MeV and at 6.20 MeV have large transition strengths. 
The state at 4.83 MeV is generated dominantly by the 2qp excitation of 
$\nu[310]1/2 \otimes \nu[440]1/2$ (4.75 MeV  for the 2qp excitation energy) 
with a weight, $X^{2}-Y^{2}$, of 0.88. 
The state at 6.20 MeV is generated predominantly by 
the 2qp excitation of $\nu[310]1/2$ and the discretized state of $\Omega^{\pi}=1/2^{+}$ (6.26 MeV) 
with a weight of 0.75 and 
slightly by the 2qp excitations of $\nu[312]3/2 \otimes \nu[411]3/2$ (6.44 MeV) 
with 0.08 and $\nu[200]1/2 \otimes \nu[310]1/2$ (7.22 MeV) with 0.03.  

We can see another prominent peak at 9.40 MeV. 
Microscopic structure of this state is summarized in Table~\ref{40Mg_0-}. 
This state is generated by many 2qp excitations as well as the $K^{\pi}=1^{-}$ 
state at 8.22 MeV representing the pygmy dipole mode. 
The contribution of the qp excitation into the non-resonant continuum state 
is larger than the pygmy $K^{\pi}=1^{-}$ state. 
The transition density to this state is shown in the upper panel of Fig.~\ref{trans_density}. 
Although the transition density represents the pygmy dipole character, 
an oscillation of the neutron skin against the isoscalar core along the symmetry axis, 
the amplitude is smaller than that of the transition density to the $K^{\pi}=1^{-}$ state at 8.22 MeV. 

\subsection{Low lying states in $^{36}$Mg and $^{38}$Mg}

Figures~\ref{Mg_dip_strength}(a), \ref{Mg_dip_strength}(b), \ref{Mg_dip_strength}(d) and 
\ref{Mg_dip_strength}(e) show the transition strengths for 
the IV dipole excitation in the lower energy region in $^{36}$Mg and $^{38}$Mg. 
For the $K^{\pi}=0^{-}$ excitations, we can see a peak at around $9-10$ MeV 
both in $^{36}$Mg and $^{38}$Mg. 
The state at 9.85 MeV in $^{36}$Mg is generated by the superposition of many 2qp excitations; 
among them the 2qp excitations of $\nu[200]1/2 \otimes \nu[330]1/2$ (8.94 MeV) 
with a weight of 0.25, $\nu[330]1/2 \otimes \nu[211]1/2$ (10.3 MeV) with 0.17, and 
$\nu[202]5/2 \otimes \nu[312]5/2$ (10.3 MeV) with 0.13 
have large contributions. 
The state at 9.69 MeV in $^{38}$Mg is also generated by many 2qp excitations. 
The 2qp excitations of $\nu[440]1/2 \otimes \nu[321]1/2$ (9.85 MeV) with 0.27, 
$\nu[330]1/2 \otimes \nu[440]1/2$ (8.83 MeV) with 0.15, and 
$\nu[200] \otimes \nu[330]1/2$ (10.4 MeV) with 0.10 
have large contributions. 
The transition densities to these states have a similar spatial structure 
to the transition density to the $K^{\pi}=0^{-}$ state at 9.40 MeV in $^{40}$Mg. 

For the $K^{\pi}=1^{-}$ excitation, we can see a peak at around $8-9$ MeV 
both in $^{36}$Mg and $^{38}$Mg. 
The state at 8.67 MeV in $^{36}$Mg is generated mainly by 
the 2qp excitations of $\nu[202]3/2 \otimes \nu[321]1/2$ (8.69 MeV) with 
a weight of 0.40 and $\nu[211]1/2 \otimes \nu[310]1/2$ (8.50 MeV) with 0.37. 
The state at 8.32 MeV in $^{38}$Mg is generated predominantly by 
the 2qp excitation of $\nu[202]3/2 \otimes \nu[321]1/2$ (8.33 MeV) with 
a weight of 0.69. 

\begin{figure}[t]
\begin{center}
\begin{tabular}{cc}
\includegraphics[scale=0.5]{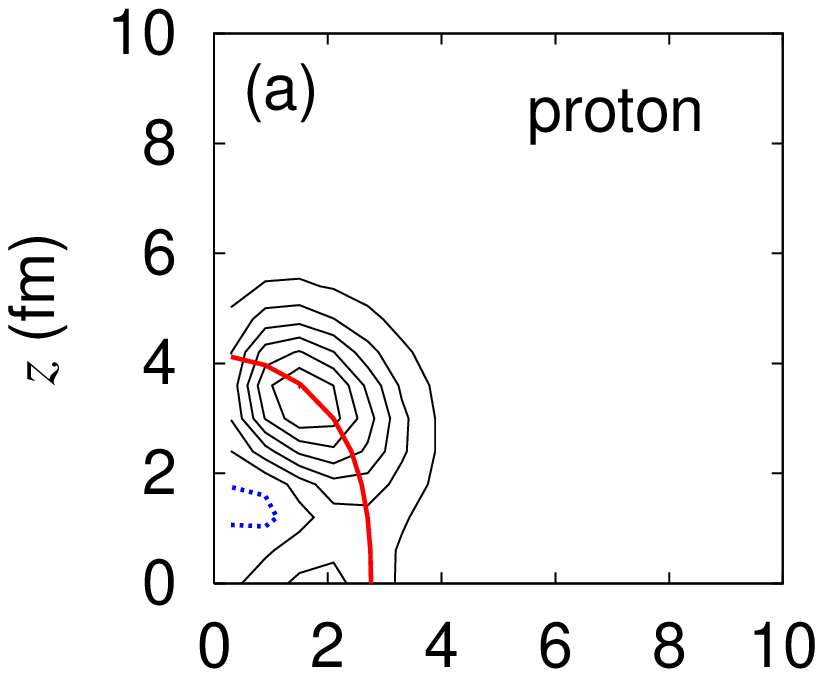}
\includegraphics[scale=0.5]{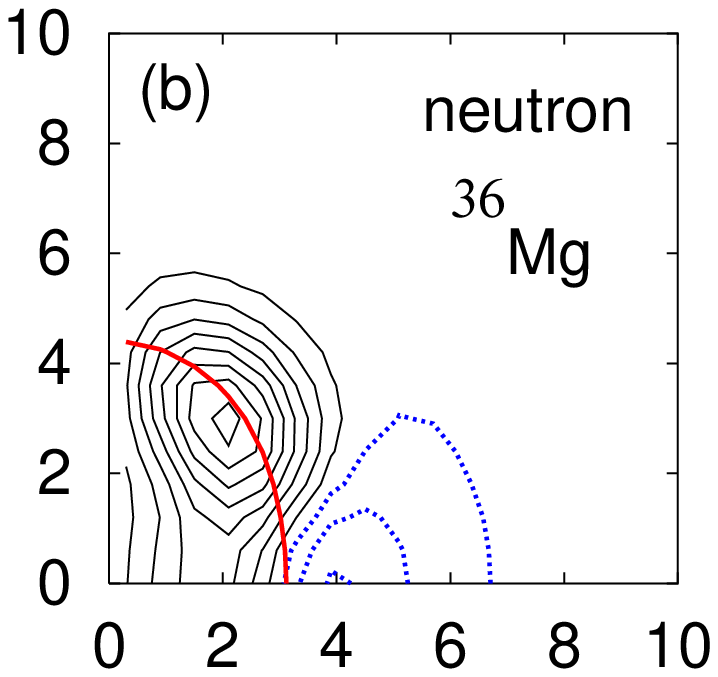} \\
\includegraphics[scale=0.5]{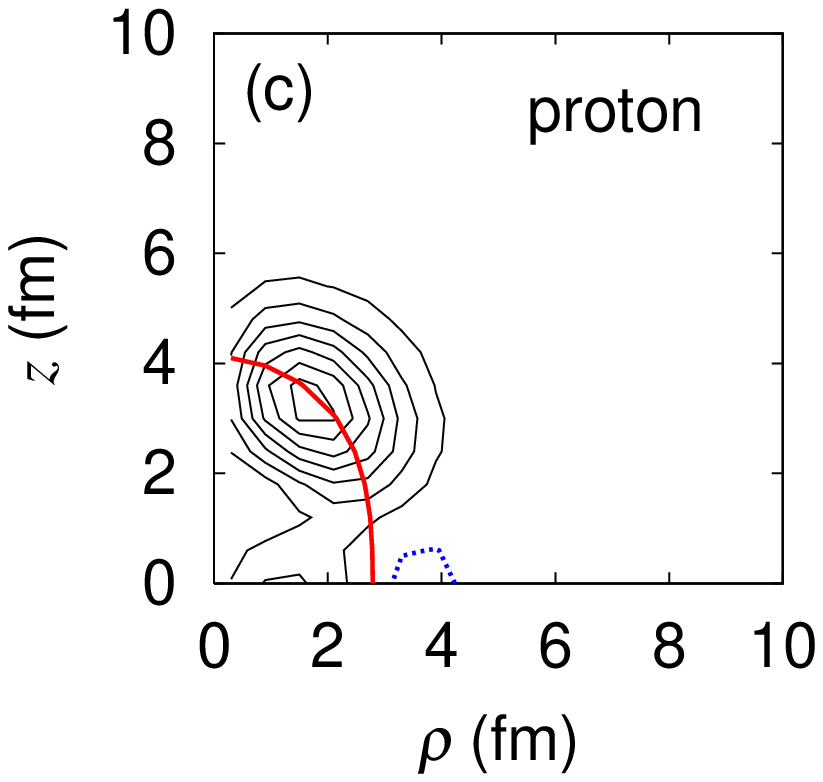}
\includegraphics[scale=0.5]{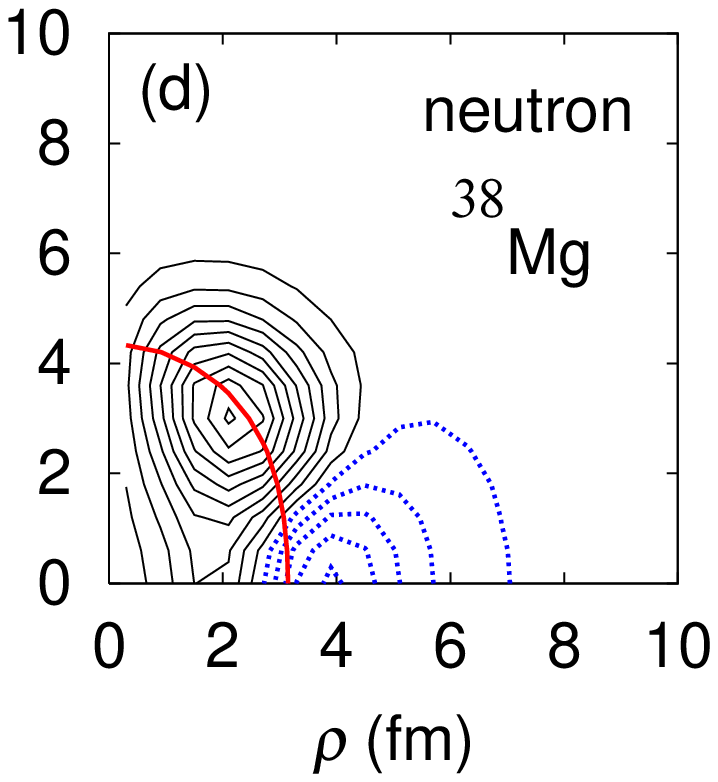}
\end{tabular}
\caption{(Color online) Same as Fig.~\ref{trans_density} but 
for the $K^{\pi}=1^{-}$ state at 8.67 MeV in $^{36}$Mg (upper) 
and the state at 8.32 MeV in $^{38}$Mg (lower). 
The neutron and proton half densities are 
0.051 fm$^{-3}$ and 0.034 fm$^{-3}$ in $^{36}$Mg and 
0.053 fm$^{-3}$ and 0.033 fm$^{-3}$ in $^{38}$Mg.
}
\label{trans_density2}
\end{center}
\end{figure}

Figure~\ref{trans_density2} shows the transition densities to the $K^{\pi}=1^{-}$ states 
in $^{36}$Mg and $^{38}$Mg. 
These states have a structure of the neutron-skin oscillation against the isoscalar core 
similarly to the pygmy state in $^{40}$Mg. 
As approaching the drip line, the neutron transition density 
has more spatially extended structure, and the amplitude is also larger 
whereas the proton transition density is unchanged. 

\section{\label{discussion}Coupling among different modes of excitation}

\begin{figure}[t]
\begin{center}
\includegraphics[scale=0.85]{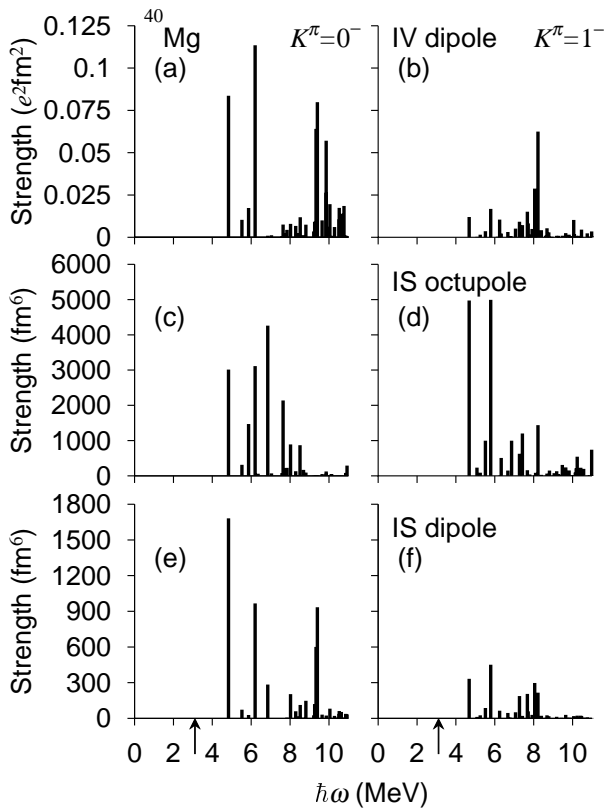}
\caption{$K^{\pi}=0^{-}$ (left) and $1^{-}$ (right) transition strengths for 
the IV dipole (upper), isoscalar (IS) octupole (middle) and IS dipole (lower) excitations 
in $^{40}$Mg.}
\label{40Mg_strength}
\end{center}
\end{figure}

In a deformed system, the angular momentum is no longer a good quantum number. 
Therefore, we could expect the coupling between the dipole and octupole modes 
of excitation in $^{40}$Mg 
as one of the unique features of the negative-parity excitation modes in a deformed system. 
Figure~\ref{40Mg_strength}(d) shows the $K^{\pi}=1^{-}$ isoscalar (IS) octupole transition strengths. 
The lowest state at 4.68 MeV and the state at 5.79 MeV have 
enhanced octupole transition strengths of 4980 fm$^{6}$ and 4990 fm$^{6}$. 
The state at 4.68 MeV is generated dominantly by the 2qp excitation of 
(i) $\nu[310]1/2 \otimes \nu[440]1/2$ (4.75 MeV) 
with a weight of 0.83, 
and the state at 5.79 MeV is generated by the 2qp excitations of 
(ii) $\nu[310]1/2 \otimes \nu[411]3/2$ (5.90 MeV) with 0.61 and 
(iii) $\nu[301]1/2 \otimes \nu[440]1/2$ (5.63 MeV) with 0.11. 
The state at 4.68 MeV has a similar structure to the $K^{\pi}=0^{-}$ state at 4.83 MeV. 
The excitation energies do not change so much with respect to 
the unperturbed 2qp excitation energies. 
Nevertheless, the transition strengths become large. 
This is because 
the unperturbed transition strengths of the 2qp excitations of (i) and (ii) 
are quite large, (i) 858 fm$^{6}$ and (ii) 2070 fm$^{6}$, 
as a consequence of the spatial extension of the quasiparticle wave functions 
around the Fermi level. 

We can see an appreciable coupling between the dipole and octupole excitations also 
for the $K^{\pi}=1^{-}$ pygmy state at 8.22 MeV 
possessing the IS octupole transition strength of 1440 fm$^{6}$. 

\begin{figure}[t]
\begin{center}
\includegraphics[scale=0.55]{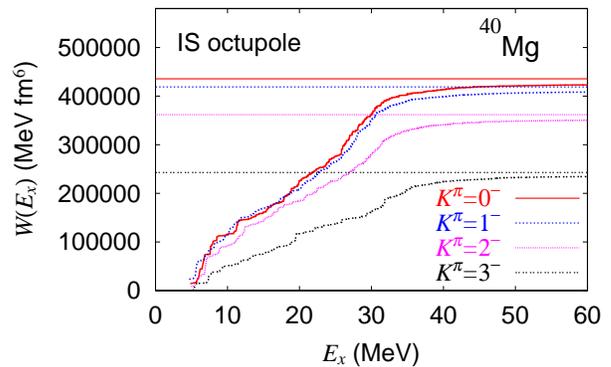}
\caption{(Color online) Same as Fig.~\ref{40Mg_EWSR} but for the IS octupole excitations.}
\label{40Mg_EWSR_oct}
\end{center}
\end{figure}

Figure~\ref{40Mg_EWSR_oct} shows the partial sum of the energy weighted
strength for the IS octupole excitations. 
The spurious component of the center-of-mass motion 
is subtracted for the EWSR values for the $K^{\pi}=0^{-}$ and $1^{-}$ excitations. 
(See Appendix for the effect of the c.m. motion on the IS octupole excitations in a deformed system.)  
The summed octupole transition strengths up to 10 MeV exhausts about $26.6 \% (27.3 \%)$ of the 
EWSR value for the IS octupole $K^{\pi}=0^{-} (1^{-})$ excitation. 
The individual eigenstates obtained in the present calculation scheme 
do not represent the collective nature. 
However, concentration of the transition strengths in the low energy region 
would be one of the unique features in drip-line nuclei. 
Investigation of this unique feature is challenging
in a more sophisticated framework that is able to handle 
the coupling to the continuum in a better way.

\begin{figure*}[t]
\begin{center}
\includegraphics[scale=0.9]{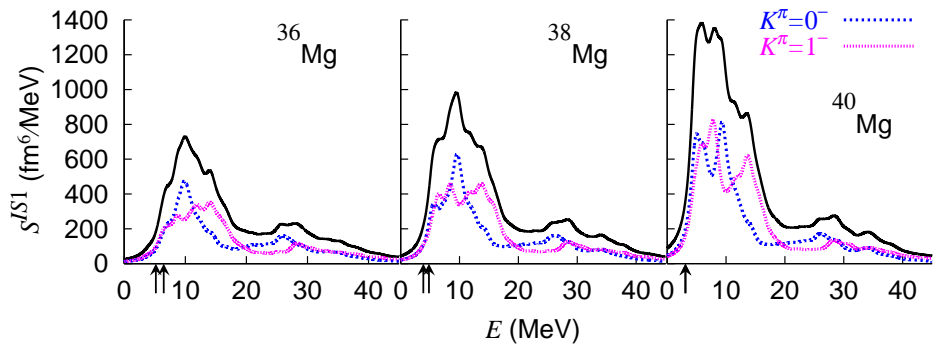}
\caption{(Color online) Same as Fig.~\ref{response} but for 
the IS compressional dipole operator (\ref{IS1}). }
\label{response2}
\end{center}
\end{figure*}

Enhancement of the transition strengths in the lower energy region 
for the IS compressional dipole excitation 
in unstable nuclei was also predicted in Ref.~\cite{ham98}. 
Since the pygmy dipole mode discussed above has both an IS and an IV structure, 
we can expect the enhancement of the strength for the IS dipole 
excitation as well as for the IV dipole excitation in Mg isotopes.

Figure~\ref{response2} shows the response functions for the IS compressional dipole mode. 
The IS dipole operator used in the present calculation is
\begin{equation}
\hat{F}_{1K}^{\mathrm{IS}}=\sum_{i \in \pi, \nu}(r^{3}_{i}-\eta r_{i})Y_{1K}(\hat{r}_{i}),
\label{IS1}
\end{equation}
where
\begin{equation}
\eta=
\begin{cases}
3\langle z^{2} \rangle +\langle \rho^{2}\rangle & (K=0) \\
\langle z^{2}\rangle +2\langle \rho^{2}\rangle & (K=\pm 1).
\end{cases}
\end{equation}
The correction factor $\eta$ originally discussed for a spherical system 
to subtract the spurious component of the c.m. motion~\cite{gia81b} 
was extended for a deformed system, 
and coincides with $\eta=\frac{5}{3}\langle r^{2} \rangle$ in the spherical limit~\cite{yos08}.

We can see a tremendous enhancement of the transition strengths above 
the neutron-emission threshold energy up to $\sim$ 15 MeV, 
where the IV dipole strengths start to have a resonance peak. 
The enhancement of the transition strengths are found 
not only for the pygmy modes but also for the less collective low-lying 
states possessing large IS octupole transition strengths 
as shown in Fig.~\ref{40Mg_strength}. 

We can see a significant coupling among the IV dipole, IS octupole and IS compressional 
dipole modes in deformed Mg isotopes close to the drip line. 
It is thus quite interesting to investigate systematically 
in a wide mass region of nuclei, where we can find the nuclear deformation, 
not only appearance of the pygmy mode 
but also coupling between the dipole and octupole excitations 
and enhancement of the transition strengths for the IS compressional dipole 
and IS octupole excitations in the low energy region. 
In neutron-rich nuclei close to the drip line, 
the low lying modes are embedded into the continuum as shown in Fig.~\ref{40Mg_strength}. 
Therefore, it is strongly desirable to develop the microscopic framework that 
enables us to investigate the continuum effects on excitation modes
and collectivity of the low lying resonance in a quantitative manner. 
Developing the QRPA on top of the HFB in Gamow basis~\cite{mic08} 
is one of the ultimate ways and has been undertaken~\cite{mic09}. 
The present work is considered to be a step toward the long-range plan 
of the microscopic and quantitative description of the collective modes 
of excitation in deformed drip-line nuclei, 
and the present results show that 
it is worthwhile to investigate the escaping widths of these states and 
collectivity of the low lying resonances in a sophisticated framework.

\section{\label{summary}Summary}
We made a detailed analysis of the low-lying dipole states in 
deformed neutron-rich Mg isotopes close to the drip line, $^{36}$Mg, $^{38}$Mg and $^{40}$Mg, 
by using the deformed QRPA employing the Skyrme SkM* and the local mixed-type pairing 
energy-density functionals. 

Above the neutron-emission threshold, we obtained several eigenstates 
having large transition strengths for the IV dipole excitation.
Because of the deformation, excitation modes with different $K$ quantum numbers 
have different excitation energies and microscopic structures. 
We obtained the collective pygmy-dipole modes at around $9 - 10$ MeV 
for the $K^{\pi}=0^{-}$ excitation 
and at around $8 - 9$ MeV for the $K^{\pi}=1^{-}$ excitation. 
These pygmy dipole modes are generated by many 2qp excitations. 

Since the pygmy dipole modes obtained in Mg isotopes have both an IS and an IV 
structure, it has been found that the transition strengths for the IS compressional 
dipole excitation are tremendously enhanced in the lower energy region. 
Furthermore, 
because of the mixing of different angular momenta in a deformed system, 
we found significant coupling among the IV dipole, IS octupole and IS compressional dipole modes 
in the low energy region. 
 
\begin{acknowledgments}  
The author thanks K.~Matsuyanagi for stimulating discussions and encouragement. 
Discussions during the RCNP workshop on the E1 $\&$ M1 excitations held in August 2009 
were useful to complete this work.
He is supported by the Special Postdoctoral Researcher Program of RIKEN. 
The numerical calculations were performed on the NEC SX-8 supercomputer
at the Yukawa Institute for Theoretical Physics, Kyoto University and 
the NEC SX-8R supercomputer at the Research Center for Nuclear Physics, Osaka University.
\end{acknowledgments}

\appendix*
\section{subtraction of the center-of-mass motion from the isoscalar octupole operator}
Because of the deformation, the $K^{\pi}=0^{-}$ and $1^{-}$ octupole excitations 
contain the spurious center-of-mass motion. 
We deal with this problem by using an operator
\begin{align}
M_{30}(\boldsymbol{r}) &\equiv r^{3}Y_{30}-\eta rY_{10} \\
&= \sqrt{\dfrac{7}{16\pi}}(2z^{3}-3z\rho^{2}-\eta^{\prime}z),
\end{align}
where $\eta^{\prime}=\sqrt{12/7}\eta$ 
for the $K^{\pi}=0^{-}$ channel 
by following the discussion in Ref.~\cite{gia81b}. 
The correction factor $\eta^{\prime}$ is determined by the condition 
of the translational invariance.

The vibrating density associated with the external field 
\begin{equation}
V(\boldsymbol{r},t)=\alpha(t)M_{30}(\boldsymbol{r})
\end{equation}
can be expressed to first order in $\alpha(t)$ by
\begin{align}
\delta \varrho(\boldsymbol{r},t) &=\alpha \nabla \cdot (\varrho_{0}\nabla M_{30}) \\
&=\sqrt{\dfrac{7}{16\pi}}\alpha
\left\{ (-6z\rho)\dfrac{\partial}{\partial \rho}
+(6z^{2}-\eta^{\prime}-3\rho^{2})\dfrac{\partial}{\partial z}
\right\}\varrho_{0}.
\end{align}
Here $\varrho_{0}$ is the ground-state density, 
and the variables $t$ and $\boldsymbol{r}$ are omitted for simplicity. 

The condition of the translational invariance of the system
\begin{equation}
\int d\boldsymbol{r} \delta\varrho rY^{*}_{10}=0 
\end{equation}
gives 
\begin{equation}
\eta^{\prime}_{K=0}= 6\langle z^{2}\rangle -3\langle \rho^{2}\rangle.
\end{equation}

The similar procedure is taken for the $K^{\pi}=1^{-}$ channel 
by using an operator
\begin{align}
M_{31}&\equiv r^{3}Y_{31}-\eta rY_{11} \\
&=-\sqrt{\dfrac{21}{64\pi}}(4z^{2}\rho-\rho^{3}-\eta^{\prime}\rho)e^{i\phi},
\end{align}
where $\eta^{\prime}=\sqrt{8/7}\eta$. It gives 
\begin{equation}
\eta^{\prime}_{K=1}=4\langle z^{2}\rangle - 2\langle \rho^{2}\rangle.
\end{equation}

In the spherical limit, the correction factors both for the $K^{\pi}=0^{-}$ and $1^{-}$ 
excitations vanish.
This is reasonable because 
the octupole excitations decouple to 
the spurious c.m. motion in a spherical system. 

The EWSR values for the IS octupole excitations 
\begin{equation}
\hat{F}^{\mathrm{IS}}_{3K}=\sum_{i \in \pi,\nu}r^{3}_{i}Y_{3K}(\hat{r}_{i})
\end{equation}
are given by
\allowdisplaybreaks[3]
\begin{align}
&\mathrm{EWSR}(\lambda=3,K=0) \notag \\ 
&=\dfrac{\hbar^{2}}{2m}A \times \dfrac{63}{16\pi}
\left( 4\langle z^{4}\rangle
+\langle \rho^{4}\rangle \right), \\ 
&\mathrm{EWSR}(\lambda=3,K=1) \notag \\ 
&=\dfrac{\hbar^{2}}{2m}A 
\times \dfrac{21}{32\pi}
\left( 16\langle z^{4}\rangle + 5\langle \rho^{4}\rangle
+ 16\langle z^{2}\rho^{2}\rangle \right), \\
&\mathrm{EWSR}(\lambda=3,K=2) \notag \\ 
&=\dfrac{\hbar^{2}}{2m}A \times \dfrac{105}{32\pi}
\left( 
8\langle \rho^{2}z^{2}\rangle
+\langle \rho^{4}\rangle \right), \\ 
&\mathrm{EWSR}(\lambda=3,K=3) \notag \\ 
&=\dfrac{\hbar^{2}}{2m}A 
\times \dfrac{315}{32\pi}\rho^{4}.
\end{align}

The EWSR values for the $K^{\pi}=0^{-}$ and $1^{-}$ excitations are corrected as
\begin{align}
&\mathrm{EWSR}^{\mathrm{cor}}(\lambda=3,K=0) \notag \\ 
&=\mathrm{EWSR}(\lambda=3,K=0) - 
\dfrac{\hbar^{2}}{2m}A \times \dfrac{7}{16\pi}\eta^{\prime 2}_{K=0}, \\ 
&\mathrm{EWSR}^{\mathrm{cor}}(\lambda=3,K=1) \notag \\
&=\mathrm{EWSR}(\lambda=3,K=1) - 
\dfrac{\hbar^{2}}{2m}A \times \dfrac{21}{32\pi}\eta^{\prime 2}_{K=1}
\end{align}
by subtracting the spurious component of the c.m. motion.

\end{document}